\documentclass{article}[12pt]
\usepackage{a4wide} 
\usepackage{graphicx}
\usepackage{amssymb}
\usepackage{algorithm}
\usepackage{algorithmic}
\usepackage{amsmath}
\usepackage{amsfonts}
\usepackage{dsfont}
\usepackage{url}
\usepackage{chemarr}
\usepackage{chemarrow}
\usepackage{bbold}
\usepackage{array} 
\usepackage[version=3]{mhchem}
\usepackage{mathabx}

\begin{document}

\title{Molecular communication networks with general molecular circuit receivers}    
\author{
Chun Tung Chou\\
      School of Computer Science and Engineering \\ University of New South Wales, Sydney \\ NSW 2052, Australia \\
       Email: {ctchou@cse.unsw.edu.au}
}

\maketitle
\begin{abstract}
In a molecular communication network, transmitters may encode information in concentration or frequency of signalling molecules. When the signalling molecules reach the receivers, they react, via a set of chemical reactions or a molecular circuit, to produce output molecules. The counts of output molecules over time is the output signal of the receiver. The aim of this paper is to investigate the impact of different reaction types on the information transmission capacity of molecular communication networks. We realise this aim by using a general molecular circuit model. We derive general expressions of mean receiver output, and signal and noise spectra. We use these expressions to investigate the information transmission capacities of a number of molecular circuits. 
\end{abstract} 


\section{Introduction} 
\label{sec:intro} 
Molecular communication networks \cite{Akyildiz:2008vt,Hiyama:2010jf,Nakano:2012dv} consist of transmitters and receivers communicating with each other via signalling molecules. The transmitters may encode the messages in concentration or emission frequency of signalling molecules. When these signalling molecules reach the receivers, they trigger one or more chemical reactions within the receivers to enable the messages to be decoded. Natural molecular communication networks are ubiquitous in living organisms, e.g. multi-cellular organisms make extensive use of molecular communication to regulate body functions \cite{Alberts}. There is an increasing interest to understand and design {\sl synthetic} molecular communication networks in both the synthetic biology \cite{Basu:2005cq} and communication engineering communities \cite{Akyildiz:2008vt,Hiyama:2010jf,Nakano:2012dv}. Such synthetic molecular communication networks can be used as sensor networks for cancer detection and treatment \cite{Atakan:2012ej}, and many other applications \cite{Nakano:2012dv}. 

An important research problem in molecular communication networks is receiver design. We will refer to the set (or networks) of chemical reactions at the receiver as a molecular circuit. When signalling molecules arrive at a receiver, the molecular circuit produces a number of output molecules. The counts of output molecules over time is the output signal of the receiver. A few different reactions have been considered in the literature: ligand-receptor binding \cite{Pierobon:2011ve}, Michaelis-Menten \cite{Noel:2013tr} and reversible conversion \cite{Chou:rdmex_nc}. Each of these papers assumes a specific reaction type but there does not appear to be work on comparing the impact of different reaction types. The intention of this paper is to address this gap. The main contributions of this paper are:

\begin{itemize}
\item We present a general molecular circuit model to enable different reactions to be modelled.
\item We derive the mean output signal of the receiver and show how the mean output depends on the parameters of the general molecular circuit model.
\item We derive the signal and noise spectra of the receiver output signal. This allows us to characterise the noise due to diffusion and reactions. It also allows us to compare different molecular circuits in terms of their information transmission capacity.  
\end{itemize}

The rest of the paper is organised as follows. We present our model for transmission medium and transmitter in Section \ref{sec:model}. The general molecular circuit receiver model will be presented in Section \ref{sec:model:rec}. The models in Sections \ref{sec:model} and \ref{sec:model:rec} are combined in Section \ref{sec:complete} to form a complete model. We then use the complete model to derive the mean output response in Section \ref{sec:mean}, and signal and noise spectra, and information transmission capacity in Section \ref{sec:spec}. In Section  \ref{sec:num}, we use numerical examples to compare and understand the properties of a number of molecular circuits. Related work is discussed in section \ref{sec:related}. Finally, Section \ref{sec:con} concludes the paper.

\section{Modelling the transmission \\ medium and transmitters} 
\label{sec:model} 
The aim of this and the next sections is to present a model for molecular communication networks. This section focuses on the transmission medium and transmitters, while the next section focuses on the receivers. 

A molecular communication network consists of multiple transmitters and receivers. In this paper, we limit ourselves to one transmitter and one receiver. We assume the transmitter uses one type of signalling molecules $L$. Generalisation to multiple types of non-interacting signalling molecules is straightforward. 

\subsection{Transmission medium} 
\begin{figure}
\begin{center}
\includegraphics[page=1,trim=3cm 10cm 0cm 3cm ,clip=true, width=10cm]{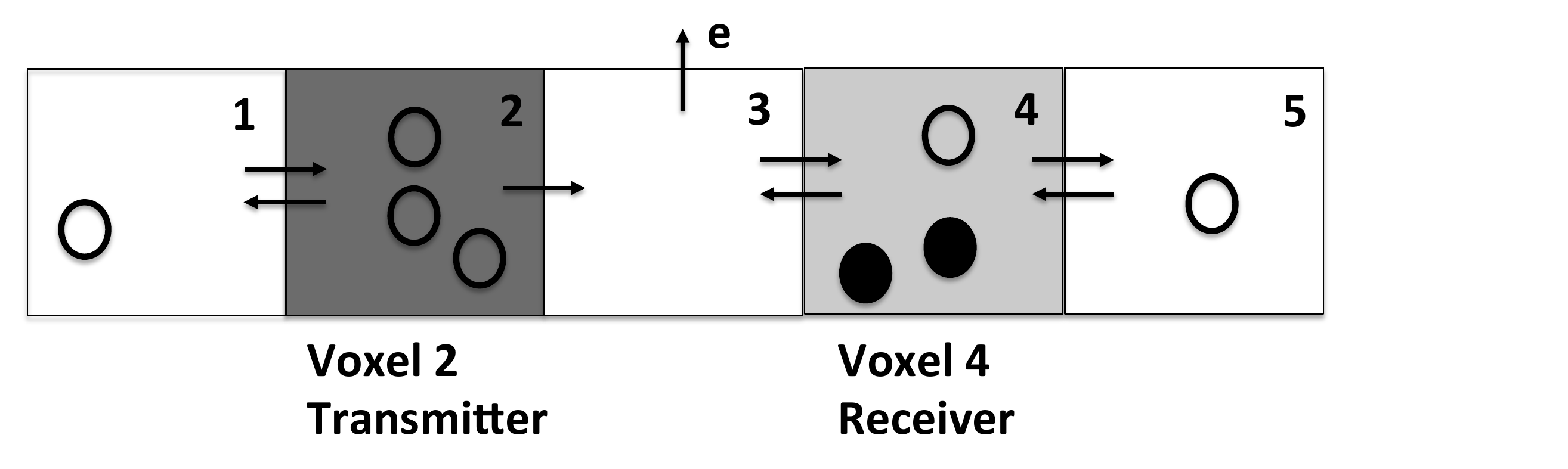}
\caption{Model of molecular communication networks. Each square is a voxel. Unfilled and filled circles represent, respectively, signalling and output molecules.}
\label{fig:model}
\end{center}
\end{figure}

We model the transmission medium as a three dimensional (3-D) space with dimensions $\ell_X \times \ell_Y \times \ell_Z$, where $\ell_X$, $\ell_Y$ and $\ell_Z$ are integral multiples of length $\Delta$. That is, there exist positive integers $N_x$, $N_y$ and $N_z$ such that $\ell_X = N_x \Delta$ and $\ell_Y = N_y \Delta$, $\ell_Z = N_z \Delta$. The 3-D volume can be partitioned into $N_x \times N_y \times N_z$ cubic {\sl voxels} of volume $\Delta^3$. Figure \ref{fig:model} shows an arrangement with $N_x = 5$ and $N_y = N_z = 1$.

We refer to a voxel by a triple $(x,y,z)$ where $x$, $y$ and $z$ are integers or by a single {\sl index} $\xi \in [1,N_x N_y N_z]$ where $\xi(x,y,z) = x + N_x (y-1) + N_x N_y (z-1)$. The indices for the voxels are shown in Figure \ref{fig:model}. 

Diffusion is modelled by molecules moving from one voxel to another. Diffusion from a voxel to a non-neighbouring voxel is always not allowed. The diffusion from a voxel to a neighbouring voxel may or may not be allowed. This can be used to specify different modelling constraints. We use a few examples in Figure \ref{fig:model} to explain this: 
\begin{enumerate}
\item For voxel 4, the diffusion of signalling molecules $L$ is allowed in both directions, i.e. in and out of the voxel. The four arrows are used to indicate this. 
\item Signalling molecules can only diffuse from voxel 2 to voxel 3, but not in the opposite direction. This may be used to model selected permeability of certain cell membranes. 
\item With the exception of the top surface of voxel 3, diffusion to the outside of the medium is not allowed. Our model can be used to capture standard boundary conditions such as reflecting and absorbing boundaries. 
\end{enumerate}

We assume that the medium is homogeneous with the diffusion coefficient for $L$ in the medium is $D$. Define $d = \frac{D}{\Delta^2}$. If a molecule is allowed to diffuse from a voxel to another, it takes place at a rate of $d$, i.e. within an infinitesimal time $\delta t$, the probability that a molecule diffuses to a neighbouring voxel is $d \delta t$. It is possible to model inhomogeneous medium in this framework, see \cite{Chou:jsac_arxiv}, but we will not consider it here. 

The rate at which the signalling molecules leave the medium is similarly defined, e.g., in Figure \ref{fig:model}, signalling molecules leave the top surface of voxel 3 (i.e. leaving the medium) at a rate of $e$. 

We assume the transmitter and the receiver each occupies a {\sl distinct} voxel. However, it is straightforward to generalise to the case where a transmitter or a receiver occupies multiple voxels. The transmitter and receiver are assumed to be located, respectively, at the voxels with indices $T$ and $R$. For example, in Figure \ref{fig:model}, voxel 2 (dark grey) contains the transmitter and voxel 4 (light grey) contains the receiver. Hence $T = 2$ and $R = 4$. 

\subsection{Transmitters} 
\label{sec:model:transmitters}
We model the transmitter by a function of time which specifies the {\em emission rate} of signalling molecules by the transmitter. We use $u(t)$ to denote the transmitter emission rate at time $t$. This means, in the time interval $[t,t+\delta t)$, the transmitter emits $u(t) \delta t$ signalling molecules. We assume $u(t)$ to be the sum of a deterministic part $c(t)$ and a random part $w(t)$, i.e. $u(t) = c(t) + w(t)$, with $w(t)$ having zero-mean. 

In molecular communication networks, a transmitter is likely to consist of a set of chemical reactions. These chemical reactions can use multiple intermediate chemical species in order to produce the signalling molecules. In this paper, we do not model the chemical reactions of the transmitter. 
We will also make two assumptions on the transmitters: (1) There is {\sl no feedback} from signalling molecules $L$ to the intermediate chemical species that produce $L$ in the transmitter; (2) The signalling molecule $L$ does not degrade in the transmitter. These two assumptions allow us to focus the analysis on the receiver and come out with clean-cut interpretation. It is our intention to remove these two assumptions in future work. We remark that the reader may appreciate more fully why these two assumptions are necessary after seeing the results in Section \ref{sec:mean} as the transmitter can be considered to be the dual of the receiver.

\subsection{Diffusion only subsystem} 
This section serves two purposes. First, we want to introduce the concept of diffusion only subsystem, a concept that we will make use of later on. Second, we want to give an example on how the medium and transmitter are modelled. 

We consider the molecular communication network in Figure \ref{fig:model} assuming that the receiver reaction mechanism has been {\sl removed}. This means that the network contains only signalling molecules and no reactions can take place. In the diffusion only system, the state of the system is the number of signalling molecules in the voxels. Let $n_{L,i}(t)$ denote the number of signalling molecules in the voxel with index $i$ at time $t$. The state $n_L(t)$ of this network is:
\begin{align}
n_L(t) =
\left[ \begin{array}{ccccc}
 n_{L,1}(t) & n_{L,2}(t) & n_{L,3}(t) & n_{L,4}(t) & n_{L,5}(t) 
\end{array} \right]^T
\end{align} 
where superscript $^T$ denotes matrix transpose. We remark that we also use $T$ and its subscripted form $_T$ to indicate the index of the transmitter voxel. Although the same symbol $T$ is used, its meaning can be deduced from its context. 

We adopt the convention that the states in $n_L(t)$ are ordered sequentially according to the voxel index. This means that the $T$-th and $R$-th state are, respectively, the number of signalling molecules in the transmitter and receiver. For example, for Figure \ref{fig:model}, $n_{L,R} = n_{L,4}$ is the number of signalling molecules in the receiver voxel. 

The state in the diffusion only subsystem can be changed by three types of events: (1) diffusion to a neighbouring voxel; (2) signalling molecule leaving the medium; and (3) emission of signalling molecules by the transmitter. We will look at each of these events in turn. 

For the diffusion to a neighbouring voxels, we take the diffusion from voxel 1 to voxel 2 as an example. This event takes place at a rate of $d n_{L,1}$ and each time this event takes place, $n_{L,1}$ is decreased by 1 and $n_{L,2}$ is increased by 1. We can model the change in the number of signalling molecules in the voxels by using the {\sl jump vector} $q_{d,1} = [-1,1,0,0,0]^T$ where the subscript $d$ is used to indicate that this jump vector comes from the diffusion only subsystem. If an instance of this event occurs, the state will jump from $n_L(t)$ to $n_L(t) + q_{d,1}$. As mentioned earlier, this event occurs at a rate of $d n_{L,1}$ and we will denote this by a {\sl jump rate function} $W_{d,1}(n_L(t))$ $(= d n_{L,1})$ to show that this rate is a function of the state. For the network in Figure \ref{fig:model}, there are 7 inter-voxel diffusion events; we will denote their jump vectors and jump rates by $q_{d,j}$ and $W_{d,j}(n_L(t))$ where $j = 1,..,7$. 

The signalling molecules in the network in Figure \ref{fig:model} can leave the medium via the top surface of voxel 3. This can be modelled by a jump vector of $q_{d,8} = [0,0,-1,0,0]$ and a jump rate function of $W_{d,8}(n_L(t)) = e n_{L,3}$.  
{
The transmitter emits $u(t) \delta t$ molecules at time $t$. We model this by adding this number of molecules to voxel $T$ ($=$ the index of the transmitter voxel) at time $t$. 

With the 8 jump vectors and jump rate functions, we can find a matrix $H$ such that $\sum_{j = 1}^8 q_{d,j} W_{d,j}(n_L(t)) = H n_L(t)$. The $H$ matrix for the network in Figure \ref{fig:model} is: 
\begin{align}
H =  
\left[ \begin{array}{ccccc}
-d & d & 0 & 0 & 0 \\
d & -2d & 0 & 0 & 0 \\
0 & d & -d-e & d & 0  \\
0 & 0 & d & -2d & d  \\
0 & 0 & 0 & d & -d  \\
\end{array} \right] 
\label{eqn:H} 
\end{align}

The dynamics of the diffusion only subsystem can be modelled by the stochastic differential equation (SDE) \cite{Gardiner}:   
\begin{align}
\dot{n}_L(t) & = H n_L(t) + \sum_{j = 1}^{J_d} q_{d,j} \sqrt{W_{d,j}(\langle n_L(t) \rangle)} \gamma_j + {\mathds 1}_T u(t) 
\label{eqn:sde:do} 
\end{align}
where $\langle n_L(t) \rangle$ denotes the mean of $n_L(t)$, $\gamma_j$ is continuous-time white noise with unit power spectral density with $\gamma_{j_1}$ independent of $\gamma_{j_2}$ for $j_1 \neq j_2$, and ${\mathds 1}_T$ is a unit vector with a 1 at the $T$-th element. The integer $J_d$ is the total number of jump vectors in the diffusion only subsystem; $J_d = 8$ for the example in Figure \ref{fig:model}. The noise  $\gamma_j$ is needed to correctly model the stochastic properties of the system. 

It is important to point out that the elements in $n_{L}(t)$, which have the interpretation of the number of molecules, is strictly speaking a {\sl discrete} random variable. The SDE is an approximation which holds when the order of the number of molecules is ${\cal O}(100)$ \cite{deRonde:2012fs}. However, as far as the first and second order moments are concerned, the SDE \eqref{eqn:sde:do} gives the same result as a master equation formulation that assumes the number molecules is discrete \cite{Warren:2006ky}.


\section{General receiver model} 
\label{sec:model:rec} 
When a signalling molecule $L$ arrives at a receiver, it may react, via one or more chemical reactions, to produce one or more {\sl output molecules} $X$. We assume that these reactions can only take place within the receiver voxel. We also assume that the output molecules cannot leave the receiver voxel. The output signal of a receiver is the counts of output molecules over time. 

We first present five different types of receiver molecular circuits in Section \ref{sec:rec:ex}. These different circuits are selected to demonstrate different interactions between the signalling and output molecules. Most of these circuits have been studied in biophysics literature \cite{Warren:2006ky,deRonde:2012fs}. The general receiver structure will be presented in Section \ref{sec:rec:gen}.

\subsection{Example receiver molecular circuits}
\label{sec:rec:ex} 
We present five example receivers. The first four examples consists of only two chemical species: signalling molecule $L$ and output molecule $X$. The last example receiver also has an intermediate chemical species $V$. We will use $\emptyset$ to denote chemical species that we are not interested in and whose quantity will not be tracked in the mathematical equations. 

The example receivers consist of 2--5 chemical reactions. For each reaction, we present the chemical formula as well as the jump vector and jump rate function. The jump rate in this case is the same as the reaction rate. The jump vectors and jump rates will be used later in a SDE model. The dimension of the jump vector is the same as the number of chemical species in the receiver. We adopt the convention that the first (reps. last) element of the jump vector shows the change in the number of signalling molecules (output molecules) in the receiver voxel. 

All the molecular reactions considered in this paper are linear. These linear reactions can be considered to be linearisation of nonlinear mass kinetic equations about an equilibrium. This is also similar to considering Linear Noise Approximation \cite{Gardiner}. We assume that all reaction rate constants have been suitably normalised with respect to the size of voxel. The reaction rates are always of the form of the product of a reaction rate constant and the number of a chemical species. 

In the following description, $n_{L,R}$, $n_X$ and $n_V$  denote, respectively, the number of signalling molecules in the receiver voxel, output molecules and intermediate species. The symbols $k_+$, $k_-$ and $k_i$ ($i = 0,...,5$) denote reaction rate constants. Each reaction will be described by its chemical formula (on the left-hand side), and jump vector and jump rate (on the right-hand side). The five example receivers are: 

\begin{enumerate} 
\item The {\em reversible conversion} (RC) receiver has 2 reactions: 
\begin{align}
L &  \rightarrow X, 		& \left[ \begin{array}{cc} -1 & 1 \end{array} \right]^T&, {k}_{+} n_{L,R}  \label{cr:rc1}  \\
X & \rightarrow L, 		& \left[ \begin{array}{cc} 1 & -1 \end{array} \right]^T&, {k}_{-} n_X       \label{cr:rc2}
\end{align}
In the forward reaction \eqref{cr:rc1}, signalling molecules $L$ are converted to output molecules $X$ at a jump rate (or reaction rate) of ${k}_{+} n_{L,R}$. The jump vector shows the change in the number of $L$ and $X$ molecules. If a reaction \eqref{cr:rc1} occurs, one molecule of $L$ is consumed to produce one molecule of $X$, and this is indicated by the jump vector in \eqref{cr:rc1}. The reverse reaction in \eqref{cr:rc2} can be similarly interpreted. 

\item The {\em conversion plus degradation} (CD) receiver has 2 reactions: 
\begin{align}
L &  \rightarrow X, 		& \left[ \begin{array}{cc} -1 & 1 \end{array} \right]^T&, {k}_{+} n_{L,R}  \label{cr:cd1}  \\
X & \rightarrow \emptyset, & \left[ \begin{array}{cc} 0 & -1 \end{array} \right]^T&, {k}_{-} n_X       \label{cr:cd2}
\end{align}
The forward reaction \eqref{cr:cd1} converts signalling molecules $L$ into output molecules $X$, in the same way as \eqref{cr:rc1}. The output molecule $X$ degrades at a rate of ${k}_{-} n_X$. Note that the jump vector for reaction \eqref{cr:cd2} says that each time this reaction occurs, the number of output molecules is reduced by one. 

\item The {\sl linear catalytic} (CAT) receiver consists of two reactions: 
\begin{align}
L &  \rightarrow L + X, 	&  \left[ \begin{array}{cc} 0 & 1 \end{array} \right]^T&, {k}_{+} n_{L,R}  \label{cr:cat1}  \\
X & \rightarrow \emptyset, &  \left[ \begin{array}{cc} 0 & -1 \end{array} \right]^T&, {k}_{-} n_X       \label{cr:cat2}
\end{align}
In \eqref{cr:cat1}, the signalling molecule $L$ acts as a catalyst to produce the output molecule $X$ at a rate of ${k}_{+} n_{L,R}$. Note that in \eqref{cr:cat1}, the number of signalling molecules remains unchanged before and after the reaction. This is indicated by the jump vector in \eqref{cr:cat1}, which says that every time when this reaction occurs, the number of signalling molecule remains unchanged and the number of output molecules is increased by one. Reaction \eqref{cr:cat2} is a degradation reaction similar to \eqref{cr:cd2}. 

\item The {\em catalysis plus regulation} (CATREG) receiver consists of 3 reactions: 
\begin{align}
L &  \rightarrow L + X, 	&  \left[ \begin{array}{cc} 0 &1 \end{array} \right]^T&, {k}_{+} n_{L,R}  \label{cr:catreg1}  \\
X & \rightarrow \emptyset, &  \left[ \begin{array}{cc} 0 & -1 \end{array} \right]^T&, {k}_{-} n_X       \label{cr:catreg2} \\
L & \rightarrow_{X} \emptyset, &  \left[ \begin{array}{cc} -1 & 0 \end{array} \right]^T&, k_0 n_X      \label{cr:catreg3}
\end{align}
Reactions \eqref{cr:catreg1} and \eqref{cr:catreg2} are identical to those in CAT. In reaction \eqref{cr:catreg3}, the degradation of signalling molecules $L$ in the receiver voxel is driven by the presence of the output molecules $X$ at a rate of $k_0 n_X$. This is an example of negative regulation or feedback. Note that we use $\rightarrow_{X}$ to indicate that the degradation is driven by $X$; note also that no $X$ molecules is consumed in the degradation of $L$. One may also think of \eqref{cr:catreg3} as $X + L \rightarrow X + \emptyset$. 

\item The {\em incoherent feedforward} (IFF) receiver consists of 5 reactions: 
\begin{align}
L &  \rightarrow L + X, 	&  \left[ \begin{array}{ccc} 0 & 0 & 1 \end{array} \right]^T&, {k}_{1} n_{L,R}  \label{cr:iff1}  \\
L &  \rightarrow L + V, 	&  \left[ \begin{array}{ccc} 0 & 1 & 0 \end{array} \right]^T&, {k}_{2} n_{L,R}  \label{cr:iff2}  \\
X & \rightarrow_{V} \emptyset, &  \left[ \begin{array}{ccc} 0 & 0 & -1 \end{array} \right]^T&, {k}_{3} n_{V}       \label{cr:iff3} \\
V & \rightarrow \emptyset, &  \left[ \begin{array}{ccc} 0 & -1 & 0 \end{array} \right]^T&, k_4 n_V      \label{cr:iff4} \\
X & \rightarrow \emptyset, &  \left[ \begin{array}{ccc} 0 & 0 & -1 \end{array} \right]^T&, k_5 n_X      \label{cr:iff5}
\end{align}
Reactions \eqref{cr:iff1} and \eqref{cr:iff2} are linear catalytic reactions similar to \eqref{cr:cat1}. Reaction \eqref{cr:iff3} is a negative regulation, similar to \eqref{cr:catreg3}. Reactions \eqref{cr:iff4} and \eqref{cr:iff5} are degradation reactions. Note that each jump vector consists of 3 elements, showing the change in the number of $L$, $V$ and $X$. This receiver is incoherent because the two reaction pathways $L \rightarrow V \rightarrow X$ and $L \rightarrow X$ have opposite effects on $X$. The former decreases the number of $X$ while the latter increases. 
\end{enumerate} 

The RC, RD, CAT and IFF reaction types have been studied in biophysics literature \cite{Warren:2006ky,deRonde:2012fs}. The reactions RC, RD, CAT and CATREG have been chosen to cover the possibilities: (1) $L$ is consumed or not, and (2) $X$ reverts to or interacts with $L$; see the first three columns of Table \ref{tab:cf:rec}. IFF is chosen as a representative of a more complicated molecular circuit; its property is also interesting, see Section \ref{sec:num}. 

\begin{table}
\centering
\begin{tabular}{|l|p{1.6cm}|p{1.6cm}||c|} \hline 
Receivers		& $L$ is consumed   & $X$ reverts or interacts with $L$ & ${\cal R}$ matrix \\ \hline
RC			& yes 			& yes	& $\left[ \begin{array}{c|c} -k_+ & k_-  \\ \hline k_+ & k_- \end{array} \right]$	\\[2ex] \hline
CD			& yes 			& no		& $\left[ \begin{array}{c|c} -k_+ & 0  \\ \hline k_+ & k_- \end{array} \right]$ \\[2ex] \hline
CAT			& no				& no		& $\left[ \begin{array}{c|c} 0 & 0  \\ \hline k_+ & k_- \end{array} \right]$ \\[2ex] \hline
CATREG		& no				& yes 	& $\left[ \begin{array}{c|c} 0 & k_0  \\ \hline k_+ & k_- \end{array} \right]$ \\[2ex] \hline
\end{tabular}
\caption{Classification of receivers (middle two columns). ${\cal R}$ matrix of the receivers (last column).}
\label{tab:cf:rec} 
\end{table}

\subsection{Receiver only subsystem}
\label{sec:rec:gen} 
In this section, we will write down the SDE governing the dynamics of a general receiver. We do not consider diffusion in this section. We will combine diffusion and receiver subsystems in Section \ref{sec:complete}. 

A general receiver consists of at least two chemical species: signalling molecule $L$ and output molecule $X$, but it may also contain a number of intermediate chemical species $V_1$, \ldots, etc. An example receiver with an intermediate species is IFF. We define the state of the receiver only subsystem as the number of signalling molecules in the receiver $n_{L,R}$, the number of each of the intermediate species $n_{V,i}$ and the number of output molecules $n_X$. We arrange the state so that the first and last element of the state vector, are, respectively, $n_{L,R}$ and $n_X$. The state vector $\tilde{n}_R$ of the receiver only subsystem is: 
\begin{align}
\tilde{n}_R(t) & =
\left[ \begin{array}{c|ccc}
 n_{L,R}(t) & n_{V,1}(t) & \cdots & n_{X}(t) \end{array} \right]^T 
\end{align} 

A receiver is specified by its jump vectors $q_{r,j}$ and jump rates $W_{r,j}(\tilde{n}_R(t))$ of its constituent reactions. Note the subscript $r$ is used to indicate that these parameters come from the receiver only subsystem. The jump vectors and jump rates of 5 example receivers are presented earlier. Given these jump vectors and jump rates, the dynamics of the receiver only subsystem is governed by the SDE:
\begin{align}
\dot{\tilde{n}}_R(t) & = {\cal R} \tilde{n}_R(t) + \sum_{j = J_d+1}^{J_d + J_r} q_{r,j} \sqrt{W_{r,j}(\langle \tilde{n}_R(t) \rangle)} \gamma_j 
\label{eqn:sde:ro} 
\end{align}
where $\gamma_j$ is white noise. The number of reactions in the receiver is $J_r$, e.g. $J_r = 5$ for IFF. Note that we index the reactions from $J_d+1$ from $J_d + J_r$ in preparation of combining the diffusion only and receiver only subsystems later on. The matrix ${\cal R}$ has the property ${\cal R} \tilde{n}_R(t) = \sum_{j = J_d+1}^{J_d+J_r} q_{r,j} W_{r,j}( \tilde{n}_R(t) )$. The ${\cal R}$ matrix for RC, CD, CAT and CATREG receivers are shown in Table \ref{tab:cf:rec}. The ${\cal R}$ matrix for the IFF receiver is:
\begin{align}
\left[ \begin{array}{c|cc} 0 & 0 & 0  \\ \hline k_2 & -k_4 & 0 \\ k_1 & -k_3 & -k_5 \end{array} \right]
\end{align}

The matrix ${\cal R}$ has certain structure, depending on whether the signalling molecules $L$ is consumed, and, whether $X$ (or any intermediate species) reverts or interacts with $L$. We partition $\tilde{n}_R(t)$ into two parts: 
\begin{align}
 \tilde{n}_R(t) 
 & =  \left[ \begin{array}{c|c}
 n_{L,R}(t) & n_R(t)^T  
\end{array} \right]^T \\ 
  \mbox{ where } 
n_R(t) & = 
\left[ \begin{array}{ccc}
 n_{V,1}(t) & \cdots & n_{X}(t) \end{array} \right]^T  
\end{align} 
We partition the matrix ${\cal R}$ conformally into $2\times2$ blocks:
\begin{align}
{\cal R}
 = \left[ \begin{array}{c|c}
R_{11} & R_{12} \\ \hline R_{21} & R_{22}   
\end{array} \right]
\label{eqn:R} 
\end{align} 
where $R_{11}$ is a scalar, and in general, $R_{12}$ and $R_{21}$ are row and column vectors. The ${\cal R}$ matrices in Table \ref{tab:cf:rec} have also been partitioned accordingly. If we compare the last 3 columns of the table, we find that: (1) $R_{11}$ is non-zero (resp. zero) if signalling molecule is (reps. not) consumed by the receiver; (2) $R_{12}$ is non-zero if and only if the output molecule $X$ (or an intermediate species) reverts or interacts with the signalling molecules. We will see that this block structure plays a role in understanding the behaviour of the receiver. 

\section{The complete system}
\label{sec:complete}
In this section, we combine the diffusion only and receiver only subsystems to form a complete system consisting of the transmitter, the medium and the receiver. The reason why we developed the two subsystems separately is that the behaviour of the complete system can be expressed in terms of the interconnections of the two subsystems. We will develop the complete system using the help of Figure \ref{fig:model}. 

The only interaction between the two subsystems takes place at the receiver voxel. The reader may also have noticed that $n_{L,R}(t)$ appears in the state vectors $n_L(t)$ and $\tilde{n}_R(t)$ of the diffusion only and receiver only subsystems. 

For the network in Figure \ref{fig:model}, the diffusion only subsystem says the number of signalling molecules $n_{L,R}(t)$ in the receiver voxel $R$ (= 4) is:
\begin{align}
\dot{n}_{L,R}(t) = d n_{L,3}(t) - 2 d n_{L,R}(t) + d n_{L,5}(t) + \xi_d(t) 
\end{align}
where $\xi_d(t)$ contains the noise term. For the receiver only subsystem, $n_{L,R}(t)$ evolves according to: 
\begin{align}
\dot{n}_{L,R}(t) = R_{11} n_{L,R}(t) + R_{12} n_R(t) + \xi_r(t) 
\label{eqn:sde:ro:r1} 
\end{align}
where $\xi_r(t)$ contains the noise term; note that \eqref{eqn:sde:ro:r1} is in fact the first row of \eqref{eqn:sde:ro}.

Since diffusion and reaction can take place at the same time, when the two subsystems are connected, we have: 
\begin{align}
\dot{n}_{L,R}(t) = & d n_{L,3}(t) - 2 d n_{L,R}(t) + d n_{L,5}(t) + \nonumber \\ 
& R_{11} n_{L,R}(t) + R_{12} n_R(t) + \xi_d(t) + \xi_r(t) 
\label{eqn:sde:nlr:ex} 
\end{align}
This is analogous to reaction-diffusion equation \cite{Gardiner}. 

To write down the complete system in general, we define the state of the complete system $n(t)$ as:
\begin{align}
n(t) = 
 & \left[ \begin{array}{c|c}
 n_{L}(t)^T & n_R(t)^T  
\end{array} \right]^T
\label{eqn:state} 
\end{align}

We will also need to modify the jump vectors from the two subsystems to obtain the jump vectors for the complete model; this will be explained in a moment.  We use $q_j$ and $W_j(n(t))$ to denote the jump vectors and jump rates of the combined model. The SDE for the complete system is:
\begin{align}
\dot{n}(t) & = A n(t) + \sum_{i = 1}^{J} q_j \sqrt{W_j(\langle n(t) \rangle)} \gamma_j + {\mathds 1}_T u(t) 
\label{eqn:sde:rd} 
\end{align} 
where $J = J_d+J_r$, and the matrix $A$ has the block structure: 
\begin{align}
A = 
 & \left[ \begin{array}{c|c}
H + {\mathds 1}_R^T {\mathds 1}_R R_{11}  &   {\mathds 1}_R R_{12}  \\ \hline 
R_{21}  {\mathds 1}_R^T & R_{22} 
\end{array} \right]
\label{eqn:A} 
\end{align}
where $H$ comes from the diffusion only subsystem (Note: an example of $H$ for Figure \ref{fig:model} is in \eqref{eqn:H}.) and $R_{11}$, $R_{12}$ etc come from the receiver only subsystem. The vector ${\mathds 1}_R$ is a unit vector with a 1 at the $R$-th position; in particular, note that  ${\mathds 1}_R^T n_L(t) = n_{L,R}(t)$ which is the number of signalling molecules in the receiver voxel. Note that, the coupling between the two subsystems, as exemplified by \eqref{eqn:sde:nlr:ex}, takes place at the $R$-th row of $A$. 

We now explain how the jump vectors for the combined system are formed. Let $m_d$ and $m_r$ denote the dimension of the vectors $n_L(t)$ and $n_R(t)$. The dimension of the jump vectors $q_j$ in the complete system is $m_d+m_r$. Given jump vector $q_{d,j}$ ($j = 1,...,J_d$) from the diffusion only sub-system with dimension $m_d$, we append $m_r$ zeros to $q_{d,j}$ to obtain $q_j$. The jump vectors $q_{r,j}$ ($j = J_d+1,...,J_d+J_r$) from the receiver only subsystem has dimension $m_r+1$. To obtain $q_j$ from $q_{r,j}$, we do the following: (1) take the first element of $q_{r,j}$ and put it in the $R$-th element of $q_j$; (2) take the last $m_r$ elements of $q_{r,j}$ and put them in the last $m_r$ elements of $q_j$. Note that jump rates are unchanged when combining the subsystems. 

\section{Mean output response} 
\label{sec:mean}
In this section we derive the mean output signal, i.e. the mean number of output molecules $\langle n_X(t) \rangle$ for a given transmitter emission function $u(t)$. In particular, we derive the frequency response from $c(t)$ (which is the deterministic part of the input signal $u(t)$) to $\langle n_X(t) \rangle$. The starting point of the derivation is \eqref{eqn:sde:rd}. We take the mean on both sides of \eqref{eqn:sde:rd}, and noting $\langle \gamma_j(t) \rangle = 0$ and $\langle u(t) \rangle = c(t)$, we have:
\begin{align}
\langle \dot{n} (t) \rangle & = A \langle n(t) \rangle + {\mathds 1}_T c(t) 
\label{eqn:mean1}
\end{align} 
Note this equation can also be considered as a spatial discretisation of a reaction-diffusion partial differential equation. 
Assuming zero initial conditions $\langle n(0) \rangle = 0$, we have the Laplace transform of the mean state vector is: 
\begin{align}
\langle N \rangle(s) & = (sI - A)^{-1} {\mathds 1}_T C(s) 
\label{eqn:meanstate:lt}
\end{align} 
where $I$ denotes the identity matrix. Here we adopt the convention of using the corresponding upper case letter to denote the Laplace transform of a signal. Since the number of output molecules is the last element of the state vector, we introduce the unit vector ${\mathds 1}_X$ with the last element being `1'. The Laplace transform of the mean number of output molecules $\langle n_X(t) \rangle$ is:
\begin{align}
\langle N_X \rangle(s) & = {\mathds 1}_X \langle N \rangle(s) = \underbrace{{\mathds 1}_X (sI - A)^{-1} {\mathds 1}_T}_{\Psi(s)} C(s) 
\end{align} 
By using the block structure of $A$ in \eqref{eqn:A}, inversion formula for block matrices and the matrix inversion lemma \cite{Zhou}, we have, after some manipulations:
\begin{align}
\Psi(s) & = \frac{G_{XL}(s) H_{RT}(s)}{1 - (R_{11} + G_{LL}(s)) H_{RR}(s)} 
\label{eqn:meanX}  
\end{align} 
where 
\begin{align}
H_{RT}(s)  =  & {\mathds 1}_R^T (sI - H)^{-1} {\mathds 1}_T  \label{eqn:hrt} \\
H_{RR}(s)  =  & {\mathds 1}_R^T (sI - H)^{-1} {\mathds 1}_R \label{eqn:hrr} \\
G_{XL}(s) = &  {\mathds 1}_X^T (sI - R_{22})^{-1} R_{21} \label{eqn:gxl}  \\ 
G_{LL}(s) = &  R_{12} (sI - R_{22})^{-1} R_{21} \label{eqn:glstar}  
\end{align} 
We will first interpret the transfer functions in \eqref{eqn:hrt}--\eqref{eqn:glstar}. The transfer functions $H_{RT}(s)$ and $H_{RR}(s)$ come from the diffusion only subsystem. We first point out that $H$ (an example is in \eqref{eqn:H}), which appears in $H_{RT}(s)$ and $H_{RR}(s)$, can be interpreted as the infinitesimal generator of a Markov chain describing the diffusion of the signalling molecules. The transfer function $H_{RT}(s)$ is the Laplace transform of $h_{RT}(t) = {\mathds 1}_R^T \exp(Ht) {\mathds 1}_T$ which is the probability that a signalling molecule present in the transmitter voxel $T$ at time $0$ is found in the receiver voxel $R$ at time $t$. Similarly, $H_{RR}(s)$ is the Laplace transform of $h_{RR}(t)$ which is the probability that a signalling molecule present in the receiver voxel $R$ at time $0$ is found again in the receiver voxel $R$ at time $t$. 

The transfer functions $G_{XL}(s)$ and $G_{LL}(s)$ come from the receiver only subsystem where $R_{22}$ can be viewed as the generator of a Markov chain. The transfer function $G_{XL}(s)$ is the Laplace transform of the probability that an output molecule $X$ at time $t$ is produced by a signalling molecule $L$ at time $0$. Before interpreting $G_{LL}(s)$, we first note that $G_{LL}(s)$ is zero if and only if $R_{12}$ is zero. Therefore, $G_{LL}(s)$ is non-zero if the output molecules $X$ revert to or interact with signalling molecules $L$. This means that, there is a chance that a signalling molecule is converted to an output molecule and then reverted to a signalling molecule later on. The transfer function $G_{LL}(s)$ is the Laplace transform of the probability that a signalling molecule $L$ in the receiver at time $t$ has come from a signalling molecule $L$ in the receiver at time $0$ via the molecular circuit.

We will now interpret the $\Psi(s)$ in \eqref{eqn:meanX}. We first consider the special case that $R_{11}$ and $R_{12}$ are zero. In this case, we have $\langle N_X \rangle(s) = G_{XL}(s) H_{RT}(s) C(s)$. This means the input signal $C(s)$ is transformed by $H_{RT}(s)$ to obtain the mean number of signalling molecules in the receiver voxel $\langle N_{L,R} \rangle(s)$, which is then subsequently transformed by $G_{XL}(s)$ to obtain the mean number of output molecules $\langle N_X \rangle(s)$. This holds for the CAT receiver, which does not consume signalling molecules and the output molecule $X$ does not revert to $L$. The ligand-receptor model in \cite{Pierobon:2010vg} also has a transfer function model of the form $G_{XL}(s) H_{RT}(s) C(s)$ because the number of signalling molecules is assumed to be in excess of the number of receptors \cite{Pierobon:2011ve}. 

Another special case of \eqref{eqn:meanX} has also appeared in the literature. The mean response to RC receiver in \cite[Eq.~(28)]{Chou:rdmex_tnb} can also be obtained from \eqref{eqn:meanX}. The transfer function $\Psi(s)$ in \eqref{eqn:meanX} is therefore very general. It takes into account the consumption of signalling molecules, the interaction between output molecules and/or intermediate species with the signalling molecules, as well as the possibility that a signalling molecule may leave the receiver voxel and then return later. We can now see that the general block structure of ${\cal R}$ in \eqref{eqn:R} is useful in understanding the mean output response. Lastly, we remark that $G_{RR}(s)$ can be used to affect the performance of molecular communication network. The transform functions $G_{RR}(s)$ and $G_{RT}(s)$ are affected by the membrane selectivity of the receiver, and this can be used to influence communication performance \cite{Chou:jsac_arxiv}. 

\section{Information capacity} 
\label{sec:spec} 
The complete system \eqref{eqn:sde:rd} can be viewed as a system with input $u(t)$ (emission rate of signalling molecules by the transmitter) and output  $n_X(t)$ (number of output molecules at the transmitter). We would like to study the information capacity of this system. In order to do that, we make several assumptions: (1) We assume that the deterministic part of the input $c(t)$ is a constant $c$. The value of $c$ can be used to set the operating point of the system. (2) We consider the stationary output of  \eqref{eqn:sde:rd} subject to $u(t) = c + w(t)$ where $w(t)$ is a stationary random process. This is equivalent to considering a very long code length and $w(t)$ is used to model an encoded signal from the transmitter. 

We will now derive the stationary signal and noise spectra of the system described in \eqref{eqn:sde:rd}. The system \eqref{eqn:sde:rd} models a continuous-time linear time-invariant (LTI) stochastic system where the summation term on the right-hand side of \eqref{eqn:sde:rd} is used to account for the noise in the system due to diffusion and reactions. Let $\Phi_u(\omega)$ denote the power spectral density of input signal $u(t)$ at angular frequency $\omega$. The power spectral density $\Phi_X(\omega)$ of the output signal can be readily obtained from standard results on output response of a LTI system to a stationary input \cite{Papoulis}. We have
\begin{align}
\Phi_{X}(\omega) & = \Phi_{\eta}(\omega) + | \Psi(i \omega) |^2 \Phi_u(\omega)
\end{align}
where $\Psi(s)$ is the transfer function in \eqref{eqn:meanX} and the stationary noise spectrum $\Phi_{\eta}(\omega)$ is:
\begin{align}
\Phi_{\eta}(\omega) & = \sum_{j = 1}^{J} | {\mathds 1}_X (i \omega I - A)^{-1} q_j |^2 W_j(\langle n(\infty) \rangle) 
\label{eqn:spec:noise} 
\end{align} 
where $n(t)$ is the state of the complete system in \eqref{eqn:state}, $\langle n(\infty) \rangle$ is the mean state at time $\infty$ due to constant input $c$. Note that $\langle n(\infty) \rangle$ can be calculated from the results in Section \ref{sec:mean}. 

We can divide the noise spectrum $\Phi_{\eta}(\omega)$ as the sum of the noise due to diffusion $\Phi_{\eta,d}(\omega)$ and reactions $\Phi_{\eta,r}(\omega)$, where:
\begin{align}
\Phi_{\eta,d}(\omega) & = \sum_{j = 1}^{J_d} | {\mathds 1}_X (i \omega I - A)^{-1} q_j |^2 W_j(\langle n(\infty) \rangle) 
\label{eqn:spec:noise:d} \\
\Phi_{\eta,r}(\omega) & = \sum_{j = J_d + 1}^{J} | {\mathds 1}_X (i \omega I - A)^{-1} q_j |^2 W_j(\langle n(\infty) \rangle)
\label{eqn:spec:noise:r} 
\end{align} 

One cause of diffusion noise is the diffusion of signalling molecules between neighbouring voxels. Let $v_1$ and $v_2$ be the indices for two neighbouring voxels. The jump vector $q$ corresponding to the diffusion from voxel $v_1$ to $v_2$ has a $-1$ in the $v_1$-th position and a $1$ in the $v_2$-th position of $q$. It can be shown that:
\begin{align}
 | {\mathds 1}_X (s I - A)^{-1} q |^2 
= 
\left| \frac{G_{XL}(s) (H_{R,v_1}(s)  - H_{R,v_2}(s))}{1 - (R_{11} + G_{LL}(s)) H_{RR}(s)} \right|^2 
\end{align} 
where $H_{R,v_i}(s) = {\mathds 1}_R^T (sI - H)^{-1} {\mathds 1}_{v_i}$ for $i = 1,2$. The transfer function $H_{R,v_i}(s)$ is related to the probability that a signalling molecule $v_1$ at time $0$ ends up at the receiver at time $t$. Since $v_1$ and $v_2$ are voxels next to each other, $H_{R,v_1}(s)$  and $H_{R,v_2}(s)$ are similar, so this has the effect of diminishing the diffusion noise. Another point to note is that the noise spectrum can again be expressed as transfer functions from the diffusion only and receiver only subsystems. We will take a closer look at the noise due to reactions $\Phi_{\eta,r}(\omega)$ \eqref{eqn:spec:noise:r} for CATREG in Section \ref{sec:num}. 

If the input signal $u(t)$ is Gaussian distributed, then the output signal $n_X(t)$ is also Gaussian distributed. In this case, the mutual information $I(n_X,u)$ between $u(t)$ and $n_X(t)$ is:  
\begin{align}
I(n_X,u) = \frac{1}{2} \int \log \left( 1+\frac{ | \Psi(i \omega) |^2}{\Phi_{\eta}(\omega)} \Phi_u(\omega) \right) d\omega
\label{eqn:mi}
\end{align}
The information capacity of the system is then given by the water-filling solution to \eqref{eqn:mi} subject to power constraint on the input $u(t)$  \cite{Gallager}. The input signal $u(t)$ may have certain constraints on its spectral characteristics because it is generated by a set of chemical reactions. In this paper, we will not take these constraints into consideration and plan to address this in future work. Lastly, we remark that if the input and output are not Gaussian distributed, the capacity calculated is a lower bound of the true capacity \cite{Mitra:2001ib}. We will use this method to compare the performance of different molecular circuits in the next section.

\section{Numerical examples}
\label{sec:num} 
In this section, we present numerical examples to illustrate the properties of the five receivers discussed in Section \ref{sec:model:rec}. 

\subsection{Comparing RC, CD, CAT and CATREG}

We consider a medium of 5$\mu$m $\times$ 1.67 $\mu$m $\times$ 1 $\mu$m. We assume a voxel size of ($\frac{1}{3}$$\mu$m)$^{3}$ (i.e. $\Delta = \frac{1}{3}$ $\mu$m), creating an array of $15 \times 5 \times 3$ voxels. The transmitter and receiver are located at voxels (4,3,2) and (12,3,2). 

We assume the diffusion coefficient $D$ of the medium is 1 $\mu$m$^2$s$^{-1}$. For RC, $k_+$ varies from 1 to 10; the $k_+$ value for other receivers will be discussed below.  The value of $k_-$ for all receivers is 0.1 s$^{-1}$. The value of $k_0$ for CATREG is 0.1. These values are similar to those used in \cite{Erban:2009us} and are realistic for biological systems. We assume an absorbing boundary for the medium and the signalling molecules escape from the boundary voxel surface at a rate of $\frac{d}{20}$. 

The deterministic emission rate $c$ is chosen to be 10 molecules per second. With this deterministic input rate and a given value of $k_+$ for the RC receiver, we compute the mean steady state output of the RC receiver, which will be denoted by $\alpha$. We can view $\alpha$ as an average demand on the receiver because it is the mean number of output molecules that a receiver has to produce. We adjust the $k_+$ value for the RD, CAT and CATREG receivers so that in each case, the mean number of output molecules is $\alpha$. The above process is repeated for each value of $k_+$ for the RC receiver. This method of adjusting the parameters means that we are comparing the receivers on the basis of same deterministic emission rate $c$ and the same mean number of output molecules. 

Now we have all the parameters of all receivers. For each receiver, we can use \eqref{eqn:meanstate:lt} to compute the mean state vector $n(\infty)$, which is then used to compute the noise spectrum $\Phi_{\eta}(\omega)$ \eqref{eqn:spec:noise}. The transfer function $\Psi(s)$ can be computed from \eqref{eqn:meanX}. We then maximise the mutual information in \eqref{eqn:mi} by water-filling assuming the input power constrained to be 100 pW. This gives us the capacity for the four receiver types for a particular value of $k_+$ for the RC receiver. 


\begin{figure}[!th]
\begin{center}
\includegraphics[width=10.0cm]{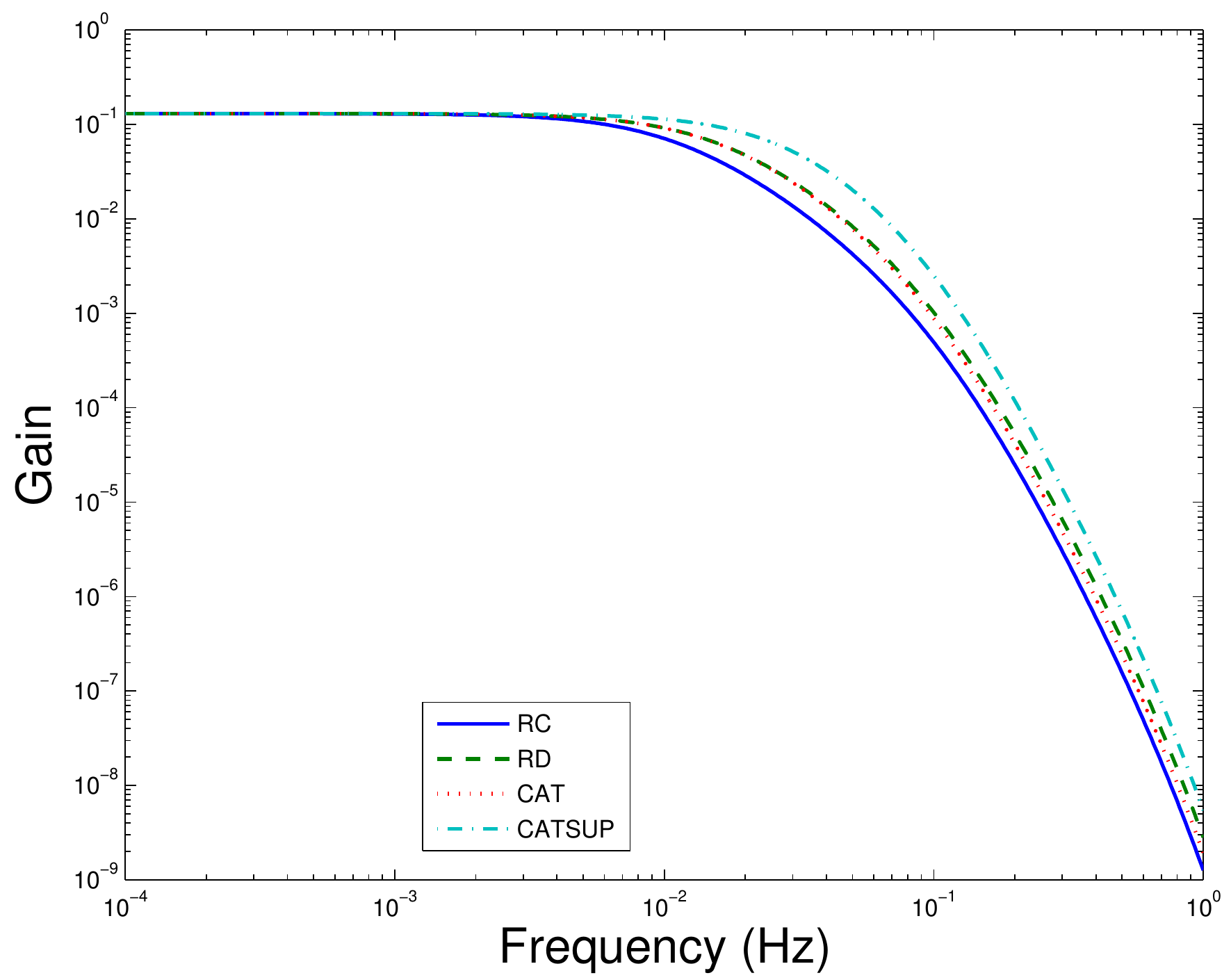}
\caption{Gain $|\Psi(i \omega)|^2 $ of RC, RD, CAT and CATREG receivers.}
\label{fig:gain}
\end{center}
\end{figure}

\begin{figure}[!th]
\begin{center}
\includegraphics[width=10.0cm]{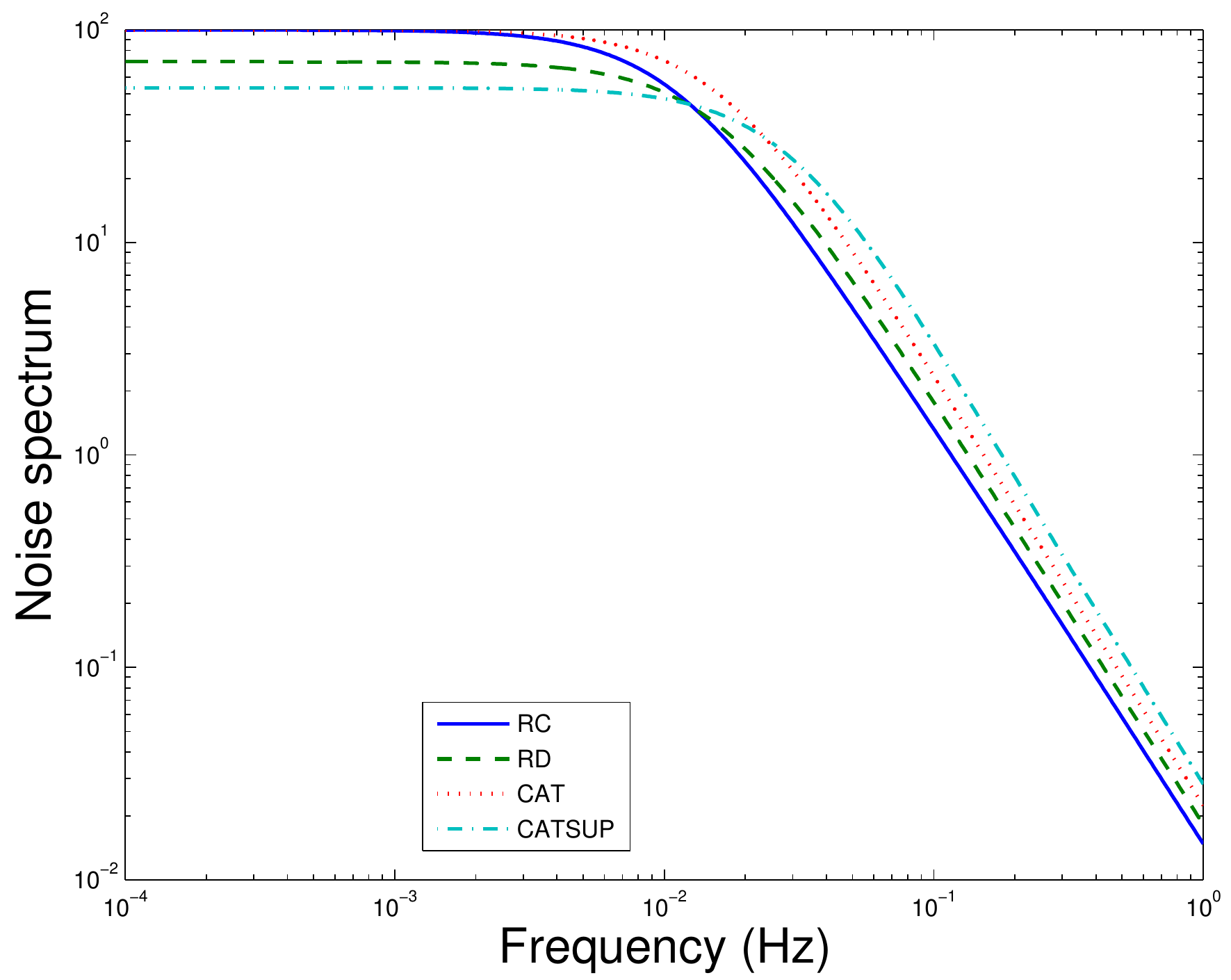}
\caption{Noise spectra of RC, RD, CAT and CATREG receivers.}
\label{fig:noise}
\end{center}
\end{figure}

Figures \ref{fig:gain} and \ref{fig:noise} show, respectively, the gain $| \Psi(\omega) |^2$ and noise spectrum $\Phi_{\eta}(\omega)$ for the four receivers. The gain spectra in Figure \ref{fig:gain} are almost the same because we have adjusted the $k_+$ values of the receivers so that they have the same mean number of output molecules. For noise spectra, CATREG has the smallest noise, followed by CD. The noise spectra for RC and CAT are similar. The difference in noise spectra is reflected in the the gain-to-noise ratio $\frac{|\Psi(i \omega)|^2}{\Phi_{\eta}(\omega)}$. CATREG has the highest gain-to-noise ratio and followed by CD. These two figures are obtained from a $k_+$ value of $10$ for the RC receiver. For small value of $k_+$, the receivers show almost the same behaviour.

%
We now vary the value of $k_+$ for the RC receivers from $1$ to $10$. For each $k_+$, we compute the capacity using water filling. The capacity of the four receivers are given in Figure \ref{fig:cap}. Receiver type CATREG has the highest capacity, followed by CD. The capacities for RC and CAT are similar. We will now take a closer look at why CATREG has a lower noise. 

\subsection{Noise in CATREG}
For the given distance of transmitter and receiver used in the calculation, the noise due to reaction in the receiver $\Phi_{\eta,r}(\omega)$ is the dominant source of noise. For the CATREG receiver, we can write $\Phi_{\eta,r}(\omega) = \Phi_{\eta,r_1}(\omega) + \Phi_{\eta,r_2}(\omega)$ where $\Phi_{\eta,r_1}(\omega)$ (resp.$\Phi_{\eta,r_2}(\omega)$) is the noise contribution due to reactions \eqref{cr:catreg1} and \eqref{cr:catreg2} (reaction \eqref{cr:catreg3}). It can be shown that 
\begin{align}
\Phi_{\eta,r_1}(\omega) & = 2 
\frac{k_- \langle n_{X}(\infty) \rangle}{\omega^2 + k_-^2} 
\left|  1 - k_0 
\Theta(i \omega) 
\right|^2  \\
\Phi_{\eta,r_2}(\omega) & = 
k_0 \langle n_{X}(\infty) \rangle \left| \Theta(i \omega)  \right|^2 
\\
\mbox{where } \Theta(s) & = 
\frac{G_{XL}(s) H_{RR}(s)}{1 - (R_{11} + G_{LL}(s)) H_{RR}(s)} 
\end{align} 
From these expressions, we see that if $\Theta(i \omega)$ has positive real part, then $k_0$ can decrease the noise in $\Phi_{\eta,r_1}(\omega)$ at the expense of increasing $\Phi_{\eta,r_2}(\omega)$. The effect of $k_0$ on $\Phi_{\eta,r_1}(\omega)$ is plotted in Figure \ref{fig:catreg:noise1}. The feedback term $k_0$ therefore has an effect of decreasing $\Phi_{\eta,r_1}(\omega)$.
The overall effect of a non-zero $k_0$ is to decrease the total noise in the receiver. Since the CAT receiver is a special case of CATREG with $k_0 = 0$, this also concludes that the noise in the CATREG receiver is smaller. 


\begin{figure}
\begin{center}
\includegraphics[width=10.0cm]{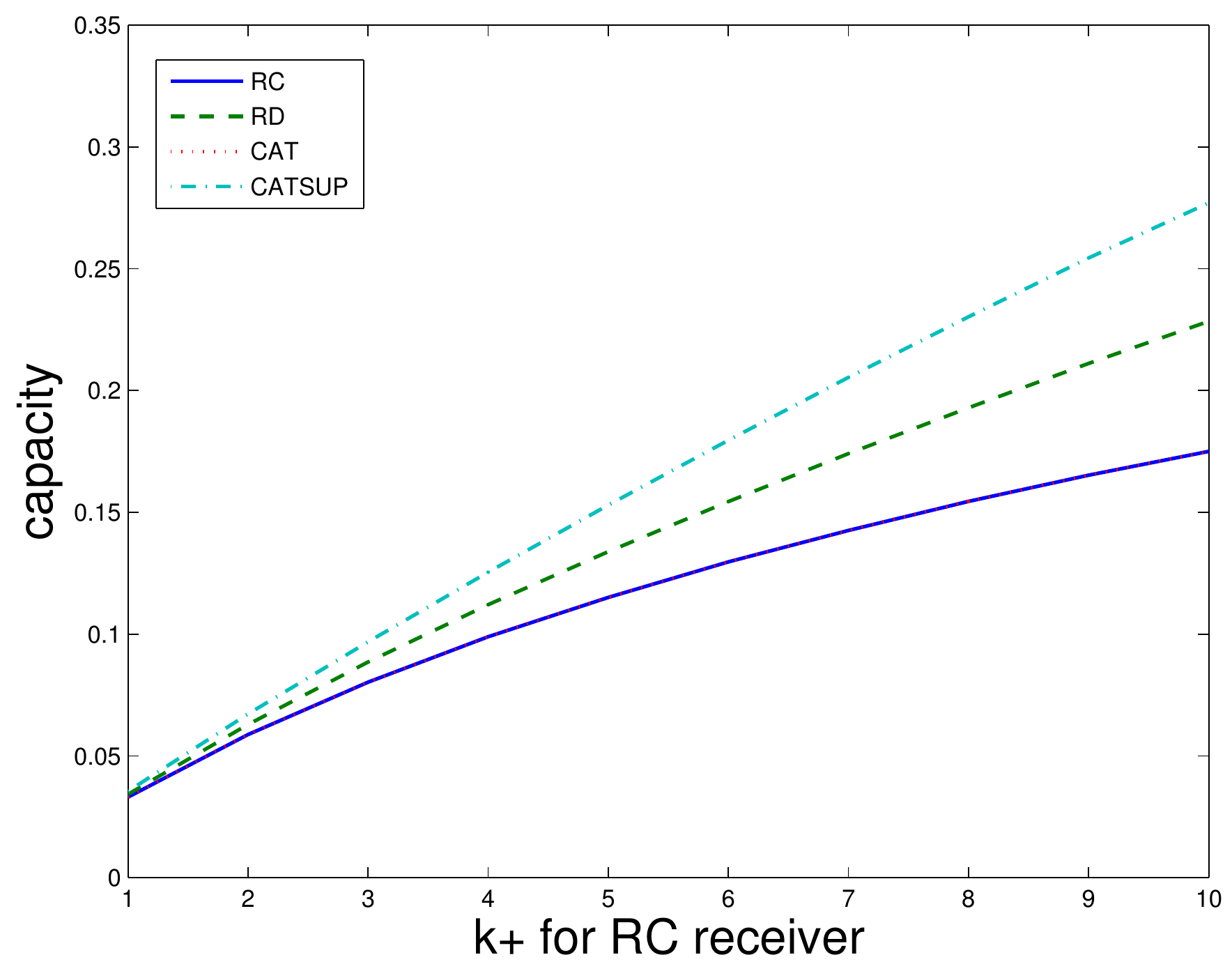}
\caption{Capacity of RC, RD, CAT and CATREG receivers.}
\label{fig:cap}
\end{center}
\end{figure}

\begin{figure}
\begin{center}
\includegraphics[width=10.0cm]{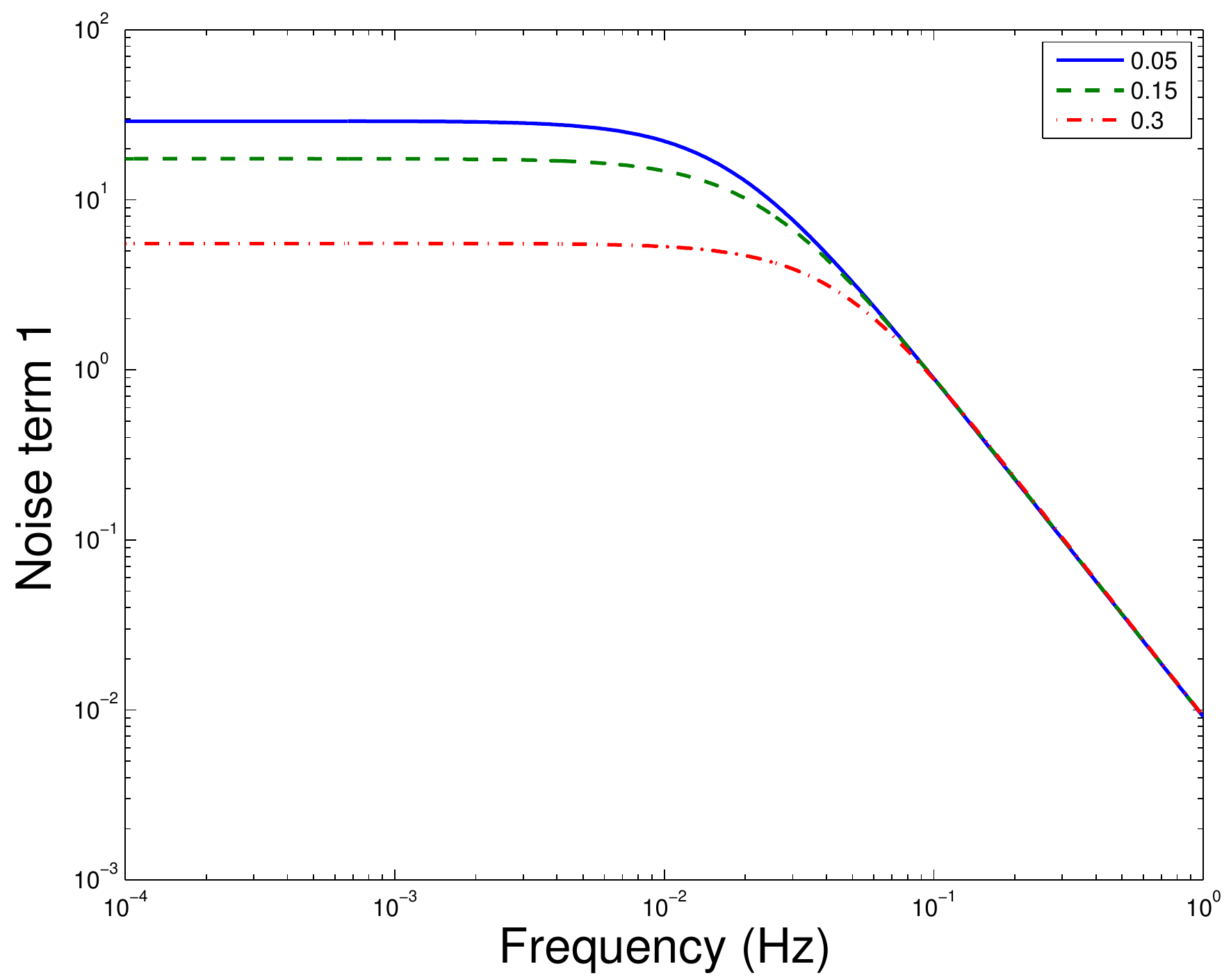}
\caption{Noise spectrum $\Phi_{\eta,r_1}(\omega)$ for CATREG receiver for different value of $k_0$.}
\label{fig:catreg:noise1}
\end{center}
\end{figure}


\begin{figure}
\begin{center}
\includegraphics[width=10.0cm]{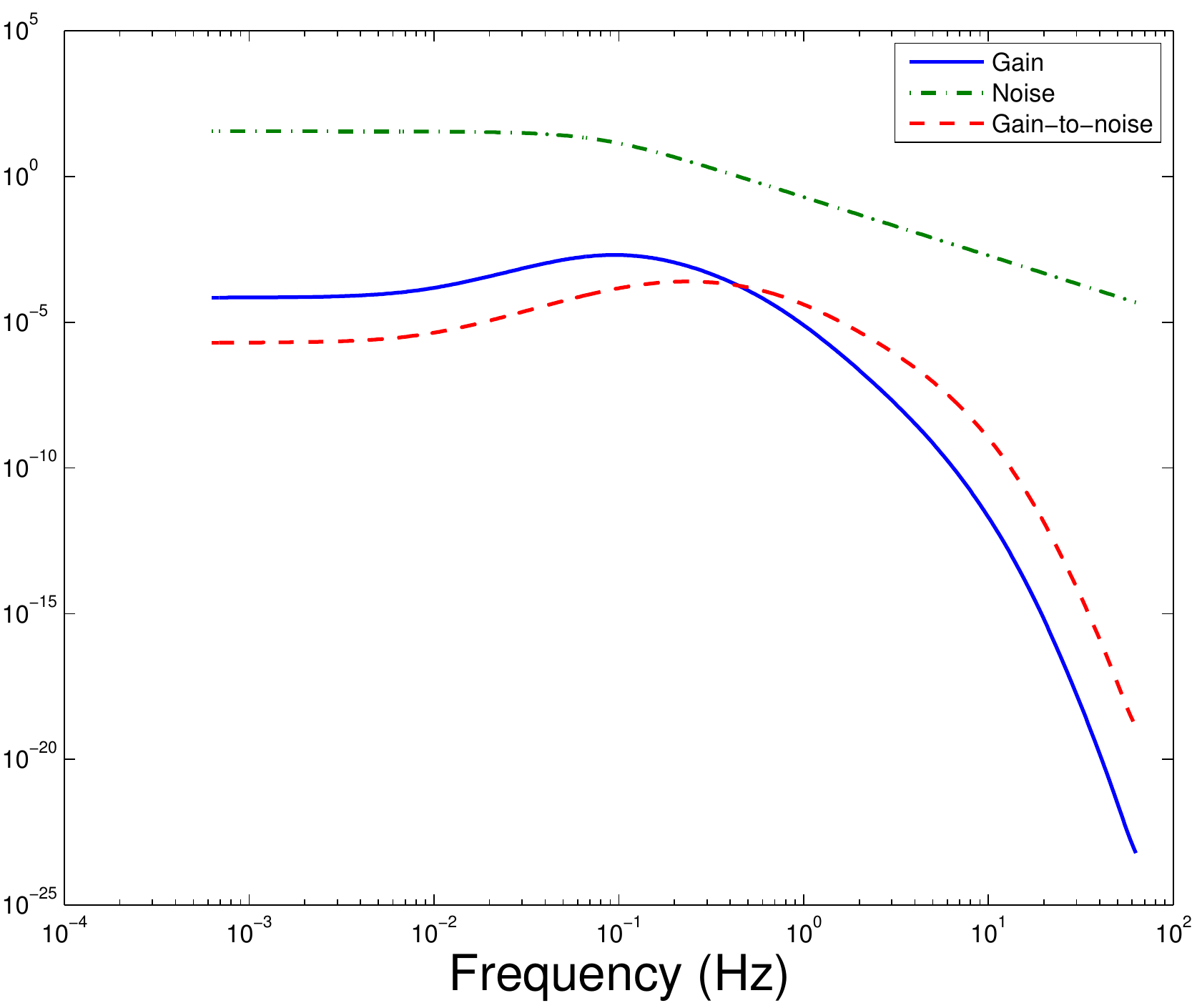}
\caption{Gain, noise spectrum and gain-to-noise of the IFF receivers.}
\label{fig:iff}
\end{center}
\end{figure}

\subsection{IFF receivers} 
We use the same transmission medium setting as before. Let $k_+ = \frac{0.1}{\Delta^3}$ and $k_- = 0.1$. The parameters of the IFF receivers are $k_1 = k_+$, $k_2 = 0.9 k_+$ and $k_3 = k_4 = k_5 = k_-$. We plot the gain, noise spectrum and gain-to-noise ratio of the IFF receiver in Figure \ref{fig:iff}. An interesting observation is that the gain has a band-pass characteristic, which is due to $G_{XL}(s)$ \eqref{eqn:gxl}. For IFF, we have
\begin{align}
G_{XL}(s) & = \frac{k_1 s + k_1 k_3 - k_2 k_4}{(s + k_3)(s+k_4)} 
\end{align}
Since $G_{XL}(\infty) = 0$, IFF does not let high frequency signals through. It is possible to find $k_i$ such that $|G_{XL}(i \omega)|$ is small at low frequencies, so suitable choice of $k_i$ can create a band-pass characteristic. We note that such receiver circuits may be suitable for decoding frequency modulated signal. 

\section{Related work} 
\label{sec:related}
Molecular communication plays a fundamental role in living organisms and has been widely studied in biology \cite{Alberts}. The study of molecular communication in the communication theory literature has been growing in the past decade. For recent review of this area, see \cite{Akyildiz:2008vt,Hiyama:2010jf,Nakano:2012dv}. Molecules in a molecular communication network can be propagated by active transport or diffusion. The former class of networks has been studied in \cite{Eckford:eq,Moore:2009eu} while the majority of the work assumes that molecules diffuse freely in the medium. This paper also assumes the transportation of molecules is by means of diffusion. 

A research problem in molecular communication networks is to understand their end-to-end performance. The authors in \cite{Pierobon:2010kz,Pierobon:2011vr,Pierobon:2011ve} investigate the mean receiver output and receiver noise assuming the receivers use ligand-receptor binding using a particle dynamics approach. The work in \cite{Chou:rdmex_tnb,Chou:rdmex_nc} derive the mean receiver output and receiver noise assuming a reversible conversion using a master equation approach. This paper proposes a general model for receiver circuit which captures the mean receiver output in \cite{Pierobon:2010kz} and \cite{Chou:rdmex_tnb} as special cases. 

Receiver design is an important topic in communication theory. There is much recent work on decoder design for molecular communication, see \cite{Noel:2013tr,Chou:2012ug,ShahMohammadian:2013jm} for example. The receiver reaction mechanisms in these papers have been chosen beforehand. In this paper, we use a general receiver model to model different reaction mechanisms. This enables us to compare the impact of different molecular circuits on the communication performance. 

The capacity of diffusion-based molecular communication network has been studied in \cite{Atakan:2010bj,Pierobon:2013cl}. Both papers consider the number of signalling molecules at the receiver as the output signal. Instead, in this paper, we use the number of output molecules of a molecular circuit as the output signal. This allows us to compare different molecular circuits. 

The biophysicists have long recognised that molecular circuits can be used to process signals. The authors in \cite{Ziv:2007wm,Tostevin:2010bo} study the signalling processing capacity of molecular circuits from an information theoretic point of view. The authors in \cite{Ma:2009wt} want to understand how the topology of the molecular circuits can impact on adaptation in chemotaxis. However, these works do not take transmitter and diffusion into consideration.

\section{Conclusions and future work} 
\label{sec:con}
This paper presents a general model for molecular communication networks. In particular, we use a receiver model which can model different types of chemical reactions. By using this general model, we derive expressions for mean receiver output, as well as signal and noise spectra. This allows us to study the information transfer capacity of different molecular circuits. We find that certain molecular circuits are able to attenuate noise better and can therefore improve molecular communication performance. In this paper, we have focused on a number of simple receiver circuits in order to focus on the generality of the model. We intend to study other molecular circuits in the future. We have made a few assumptions on the transmitters in order to focus on the performance of the receivers in this paper. We intend to remove these assumptions in future work. The models in this paper assume that the reactions are linear or the behaviour is locally linear. This is both a strength and a limitation. The strength is that we can leverage the rich theory of linear systems to understand molecular communication. The limitation is that we are not able to capture the richer types of dynamics in nonlinear systems.


\end{document}